
\input mtexsis.tex
\preprint

\pubdate{August 1992}
\pubcode{DOE-301 CPP-38}

\offparens 
\referencelist
\reference{denby1}
B.~Denby, M.~Campbell, F.~Bedeschi, N.~Chriss, C.~Bowers, and F.~Nesti,
FERMILAB-Conf-90/20
\endreference

\reference{rumel}
D.E.~Rumelhart, G.E.~Hinton, and R.J.~Williams, "Learning Internal
Representations by Error Propagation," in D.E.~Rumelhart and
J.L.~McClelland (Eds.),
\booktitle{Parallel Distributed Processing: Explorations in the
Microstructure of Cognition (Vol.1)}, MIT Press, 1986
\endreference

\reference{topnet}
H.~Baer, D.L.~Dzialo, and G.F.~Giudice, accepted for publication in
Phys. Rev. \underbar{D}
\endreference

\reference{conven}
H.~Baer, V.~Barger, and R.J.N.~Phillips, \journal Phys. Rev.;D39,3310
(1989)
\endreference

\reference{paulnbarry}
P.~Burchard and B.~Merriman, private communication
\endreference

\reference{brentpaper}
R.P.~Brent, ``Fast Training Algorithms for Multi-layer Neural Nets'',
Australian National Laboratory, submitted to Neural Networks
\endreference

\reference{numrecipes}
W.H.~Press, B.P.~Flannery, S.A.~Teukolsky, and W.T.~Vetterling,
\booktitle{Numerical Recipes in C}, Cambridge University Press,
Cambridge, England, 1988
\endreference

\reference{kdtree}
J.L.~Bentley,\journal Comm.\,ACM;23,214 (1980).
S.M.~Omohundro,\journal Complex Systems;1,273 (1987)
\endreference

\reference{dorfan}
J.~Dorfan, private communication
\endreference

\reference{pythia}
H.-U.~Bengtsson and T.~Sj\"ostrand, PYTHIA 5.5 program and manual; see
H.-U.~Bengtsson and T.~Sj\"ostrand, \journal Computer Phys.
Comm.;46,43(1987)
\endreference

\endreferencelist

\def\missEt{$\slashchar{E}_T$}
\def\xi{$x_i$}
\def\ai{$a_i$}
\def\b{$b$}
\def\xio{$x_{i_0}$}
\def\aio{$a_{i_0}$}
\def\N{$N$}
\def\xs{$x_s$}
\def\xb{$x_b$}
\def\d{$\bf d$}
\def\Q{$Q$}
\def\sig{$\Omega_S$}
\def\bac{$\Omega_B$}
\def\kd{$k$-$d$}
\def\f{$f$}
\def\mt{$m_t$}
\def\subS{${}_S$}
\def\subSpB{${}_{S+B}$}
\def\subSfrac{${}_{S/(S+B)}$}
\def\subBfrac{${}_{B/(S+B)}$}
\def\subLfrac{${}_{L/(L+R)}$}
\def\minSB{MIN\subSpB}
\def\threshSB{THRESHOLD\subSpB}
\def\maxBSB{MAX\subBfrac}
\def\minSSB{MIN\subSfrac}
\def\minLLR{MIN\subLfrac}
\def\minS{MIN\subS}
\def\lpar{\left (}
\def\rpar{\right )}
\def\half{{1 \over 2}}
\def\treetanh{\lpar {1 \over T}
				\lpar \sum_{i=1}^N a_i x_i - b \rpar
			  \rpar}
\def\weight[#1,#2,#3]{\omega^{(#1)}_{#2#3}} 
\def\neurtanh{\lpar {1 \over T}
				\sum_{i=1}^{N^{(L-1)}} \weight[L,j,i] y_i^{L-1}
			  \rpar}

\titlepage

\title
A Comparison of the Use of Binary Decision Trees and Neural Networks in Top
Quark Detection
\endtitle

\center

{\singlespaced \bf
David Bowser-Chao${}^{\hbox{\dag}}$
Debra L. Dzialo${}^{\hbox{\ddag}}$
}
\hbox{\hfil}
{\singlespaced
Center for Particle Physics
University of Texas at Austin
Austin TX 78712
}
\endcenter

\abstract
The use of neural networks for signal vs.~background discrimination in
high-energy physics experiment has been investigated and has compared favorably
with the efficiency of traditional kinematic cuts. Recent work in
top quark identification produced a neural network that, for a given top
quark mass, yielded a higher signal to background ratio in Monte Carlo
simulation than a corresponding set of conventional cuts. In this article
we discuss another pattern-recognition algorithm, the binary decision
tree. We have applied a binary decision tree to
top quark identification at the Tevatron and found
it to be comparable in performance to the neural network. Furthermore,
reservations
about the "black box" nature of neural network discriminators do not apply to
binary decision trees; a binary decision tree may be reduced to a set of
kinematic cuts subject to conventional error analysis.
\endabstract

 \vfootnote\dag{email address (Internet): davechao@bongo.cc.utexas.edu}
\vfootnote\ddag{email address (Internet): phbd064@utxvms.cc.utexas.edu}
\endtitlepage

\section{Introduction \label{sect.intro}}
Neural networks have been proposed as an adjunct to or
even replacement for cuts traditionally employed to separate
signal from background in high-energy experiments.\cite{denby1} Certainly the
development of a powerful, general training algorithm for non-recursive
neural networks\cite{rumel} has established the forward-feed,
back-propagation neural network as an important tool for pattern recognition,
both in artificial intelligence and industrial
applications.~\footnote*{See the IEEE proceedings
on neural networks of any year, for example.}
A neural network can be trained to distinguish "signal"
events from "background" events in a high-energy collider, differentiating
between the two on the basis of kinematical variables such as angular
separation, missing transverse energy \missEt, etc. One of us (D.L.D) has
investigated the use of a neural network trigger at the Tevatron
 for separating the top quark signal of one lepton plus jets
from the strong $W$-boson plus
multi-jet background.\cite{topnet} Among the results of this work was a
neural network that for a given top quark mass obtained a higher signal to
background ratio than a corresponding set of conventional cuts in Monte
Carlo simulation.\cite{conven}

There exist reservations, however, about the use of neural network triggers
in colliders. The architecture of the neural network responsible for
its success in a wide range of pattern-recognition problems
precludes straightforward error analysis, in contrast to the situation
with simple kinematic cuts. Thus we have considered another
pattern-recognition algorithm, the binary decision tree, and compared its
performance with that of the neural network in the top
quark detection problem of
\Ref{topnet}. In this particular case, the binary decision
tree does about as well
as the neural network solely on the basis of the respective increases
in signal to background. The binary decision tree,
however, may be reduced to a
set of conventional kinematic cuts, and is subject therefore to
 the usual error propagation techniques.

Section \use{sect.btree} outlines the algorithm behind the binary decision
tree,
which is essentially an automated (and optimized) search for the
series of kinematic cuts that will best isolate high signal percentage
 regions in phase space. A review of neural networks follows,
both as background for
the subsequent discussion comparing the two methods and to elucidate
difficulties encountered in the error analysis of a neural network trigger.
Finally, we examine the differences between the neural network
and binary decision tree and present the results of application
of the binary decision tree to top quark identification.

\section{Binary Decision Trees \label{sect.btree}}
In its simplest form, a signal trigger attempts to correctly
classify events as signal or background by means of a linear inequality:
$$\sum_{i=1}^{N} a_i x_i > b, \EQN singlecut$$
where the \xi\ are the kinematical variables measured for each event that
serve as input to the trigger. Given the \xi\ measured for an event,
the trigger accepts the event if,
say, the condition of \Eq{singlecut} is satisfied, and rejects it
otherwise. Of course, a single cut rarely suffices to reduce a strongly
predominant background, so that triggers usually comprise several
individual cuts administered jointly. A straightforward and common
approach to obtaining cuts is to restrict the form of \Eq{singlecut} by
setting all the \ai\ except one, \aio, to zero, so that \b\ represents
either the minimum or maximum value of \xio\ allowed for an event to be
accepted by the trigger. By plotting the distributions in \xio\ for
signal and background, \b\ can be chosen to maximize the expected signal
to background ratio of accepted events. Such cuts may be formulated for
each of the coordinates \xi\ and combined to form a set of \N\ simultaneous
conditions for event acceptance. Though one might suspect cuts of this
form to lack the power of the generalized inequality expressed by
\Eq{singlecut}, obtaining such an improved cut is rarely possible
since difficulties associated with the construction
(and interpretation) of higher
dimension plots and histograms usually limit the use of more than one
non-zero \ai\ to instances in which all non-zero \ai\ are equal (e.g., a cut
on {\it total} lepton transverse momenta $p_T$).

The aim of the binary decision tree presented here is to
enable and optimize the choice of such generalized cuts and thus
 to formulate event discriminators of higher efficiency than those
derived through standard methods. The basic algorithm is easily understood by
considering \Eq{singlecut} geometrically; the inequality defines an
\N-dimensional hyperplane that divides the phase space in two. Points on
one side of the plane are classified as signal and points on the other as
background. If the \ai\ are normalized, they define the normal to the
hyperplane, with \b\ signifying the normal distance between the plane and
origin. The centroids of the signal and background distributions,
 \xs\ and \xb, are points in this \N -dimensional space that
if not coincident define the binary decision tree's initial generalized
cut as  the hyperplane midway between \xs\ and
\xb, perpendicular to their unit separation vector \d. Thus the \ai\ are
identified with the components of \d\ and \b\ is (modulo a possible sign)
equal to half the distance between the two points.
The cut partitions phase space
into two pieces, one of which is guaranteed to have at least as high a
signal to background ratio as the parent (pre-cut) distribution.  If this
piece is deemed to have a sufficiently high expected signal to
background ratio or if further cuts would overly reduce the signal
acceptance rate, division halts. Otherwise, it is
 successively divided by generalized cuts as described above. The end
result is a series of simultaneous cuts that delimit a subregion of
accepted events. Furthermore, at each division the piece
with the lower signal to
background ratio need not be discarded; cuts may be applied to it as
well, with the aim of gaining additional pockets of signal. The hierarchy
of cuts and the partition of phase space into "signal" and "background"
regions make up the definition of the binary decision
tree.~\footnote\ddag{The binary decision tree presented here
is based on both the work of P.~Burchard with B.~Merriman (\Ref{paulnbarry})
and that of R.P.~Brent. With the exception of the scaling factor $K$, the
algorithm for optimizing hyperplanes is defined in \Ref{brentpaper}}
The recursive selection of hyperplanes (i.e., determination of
a set of \ai\ and \b) is denoted as the "training"
of the binary decision tree,
which is binary because at each step intermediate regions (nodes)
are divided into two branches. The terminal regions are the "leaves" of
the binary decision tree, and are classified as signal or background
according to the percentage of signal present. The hyperplanes thus
defined may be used exactly as the simpler conventional cuts to
implement an equivalent trigger.

In practice, the signal and background distribution functions of
\xi\ are represented efficiently (but with limited accuracy) by
sets of signal and background events, respectively \sig\ and \bac, generated
by Monte Carlo simulation. Note that while quantities
such as the percentage of signal events on one side of a hyperplane are
estimated by a count of events in \sig\ and \bac\ that fall
 on that side, even if one had an overall
signal to background ratio of $10^{-6}$ we would {\it not}
require $10^6$ times
more background than signal events for training. Given an equal number
of signal and background events, we could simply scale any count
 of background events by the factor of $10^6$.

There are limitations, however, that arise from
working with finite-size "training" sets; a hyperplane as constructed above may
be considered unsatisfactory for a number of reasons, all related to
expected statistical error present in the training distributions. The
hyperplane, for example, might leave one side with an apparently high
signal percentage but so few signal events as to render such a cut unreliable.
For this reason it is prudent to consider other candidate hyperplanes,
such as the simplified hyperplanes corresponding to traditional cuts that
lie parallel to all but one axis. A cost function \Q, used to evaluate the
desirability and/or reliability of a cut, is required to decide
among the candidate hyperplanes. We employed two different cost
functions, one of which, drawn from \Ref{brentpaper}, measures the
"entropy" produced by a candidate hyperplane and is given by:
$$Q(S_l, B_l, S_r, B_r) =
	-\log\left( {S_l!\, B_l!\, S_r!\,B_r!}
					\over
				{(S_l + B_l)!\,\,(S_r + B_r)!}
		  \right),
\EQN brentcost$$
where $S_l$ and $B_r$ are, respectively, the (possibly scaled) number of
signal events on the side of the hyperplane arbitrarily designated as
"left," and the (possibly scaled) number of background events on the other
side.

\Q\ is an implicit function of the hyperplane coordinates
$\{a_i,b\}$ through $\{S_l,B_l,S_r,B_r\}$. For an ideal hyperplane in
which $S_r = B_l = 0$, \Q\ takes on its minimum value of zero. By
selecting the hyperplane in a given set of candidate
hyperplanes with the smallest value of \Q, the current subregion
of phase space is divided into two  branches, each with a signal fraction
as far away from 0.5  as possible. Note that maximizing the difference
between the signal fractions and 0.5 does not necessarily correspond to
maximizing the signal fractions themselves. To bias the binary decision tree
more towards the latter strategy, we alternatively used the
following cost function:
$$Q(S_l, B_l, S_r, B_r) = 2
	- {\left( S_l \over {S_l+B_l} \right)}^n
	- {\left( S_r \over {S_r+B_r} \right)}^n.
\EQN ourcost$$
The binary decision trees discussed in Section \use{sect.results} have $n$ set
equal to 2, which produces a bias toward hyperplanes which create
 one branch with a higher signal to background ratio.

Both cost functions require additional constraints to prevent creation of
leaves that, although apparently high in signal percentage, have so few
events as to render them statistically meaningless. We implement these
constraints by substituting the true value of \Q\ for unacceptable
hyperplanes with a large positive constant, which thus leads to their
rejection.

Aside from its use in selecting between candidate hyperplanes,
the cost function \Q\ makes possible the {\it optimization} of a candidate
hyperplane. Rotating and translating the hyperplane, by modifying the \ai\
and \b\ respectively, shifts events from one side of the
hyperplane to  the other, increasing or decreasing $S_l$, $B_r$, etc., and
thus $Q=Q(S_l, B_l, S_r, B_r)$ as well. Continuous or discrete
optimization may be carried out to minimize \Q\ by appropriately adjusting
\ai\ and \b. If the training sets are sufficiently large to permit
interpolation
 of sorts, continuous
optimization is preferable to a discrete algorithm. Following the prescription
of \Ref{brentpaper}, $Q$ can be transformed into an analytic function
of the hyperplane coordinates $\{a_i,b\}$ by means of the following
approximations of $S_l$ and $B_r$:
$$
\EQNalign{
	S_l &= S_l(a_i, b) \cr
		&= \sum_{x_i \in \Omega_S} \half
			\left[ 1 + \tanh{\treetanh} \right] \cr
	B_r &= S_r(a_i, b) \cr
		&= \sum_{x_i \in \Omega_B} \half
			\left[ 1 - \tanh{\treetanh} \right], \EQN approx \cr}
$$
with analogous subtitutions made for $S_r$ and $B_l$. To simulate the
pre-cut signal to background ratio for training purposes, we furthermore
modified \Eq{approx} by $B_l \to K B_l$, $B_r \to K B_r$, where the
scaling factor $K$ is
$$
	K =
		{\lpar S \over B \rpar}_{\rm train}
		{\lpar S \over B \rpar}_{\rm actual}^{-1}.
		\EQN scale
$$
The first ratio is simply that of the number of signal events
used in training to that of background events, and the second is the
theoretically or experimentally known value of the pre-cut signal to
background ratio. As the temperature parameter $T \to 0$, the first line
of \Eq{approx} estimates the number of events in the signal set \sig\ that
fall on the side arbitrarily designated as the "left",
while the second line with an oppositely signed $\tanh$ term
(and multiplied by the factor $K$) gives an approximate {\it
scaled} count of background events on the right side.
With these differentiable approximations of
$S_l$, $B_l$, $S_r$, and $B_r$, \Q\ itself becomes a differentiable
function of \ai\ and \b. To minimize \Q\ and simultaneously optimize the
hyperplane, we employed the Polak-Ribiere algorithm for conjugate gradient
minimization.\cite{numrecipes} As with the exact form of \Q,
to take into account the limited size and accuracy of the training sets,
we substituted the value of \Q\ computed
 from  \Eqs{brentcost} or \Ep{ourcost} with a large positive constant if, for
example, a candidate hyperplane would leave either created branch with
less than a user-defined minimum of events.

Finally, we remark that by restricting hyperplanes to lie along
coordinate axes as in simplified cuts (by requiring all \ai\ but one,
\aio, to be equal to zero) and by replacing the $(N+1)$-dimensional
optimization of $\{a_i, b\}$ with a line optimization of \b\ alone, the
resulting algorithm is that of the \kd\ tree.\cite{kdtree} This type of
decision tree has in fact already obtained successful results in high-energy
physics, having been employed in Mark~II and Mark~III to discriminate
between electrons and pions.\cite{dorfan} The \kd\ tree demands less
cpu time for training than the binary decision tree discussed in
Section~\use{sect.results} at the cost of a generally larger number of
hyperplanes required for comparable performance.

\section{Neural Networks \label{sect.neural}}

This Section provides only the detail necessary to give some perspective
on the differences between  neural networks, binary decision trees, and
more conventional methods of separating signal from background.
A more complete introduction to neural networks may be found in \Ref{topnet}
and the references therein.
Consideration here is limited to the neural network architecture/training
method most commonly used in pattern recognition with supervised training,
the forward-feed back-propagation neural network.

The neural network is parameterized by a set of weights $\weight[L,j,i]$
that connect the nodes $y_j$ of layer $L$ with the nodes $y_i$ of the
preceeding layer $L-1$. Each  training event is assigned a numerical
classification according to background=$0$,
signal$=1$. Training the neural network consists of a
gradient descent optimization of the
weights to minimize the squared difference of the classification of
each event and the neural network function \f\ evaluated at the event's
coordinates \xi. The sum of this  ``quadratic error'' over
the training sets \sig\ and \bac\ is the network equivalent of the cost
function \Q\ used by the binary decision tree.
 The neural network function $f$ for an architecture of $M$
layers and $N^{(L)}$ nodes in layer $L$ is given as
$$
	f(x_i) = y^{(M)} \lpar y^{(M-1)} \rpar, \EQN neurfunc
$$
with the functions $y^{(L)}$ defined recursively,
$$
	y_j^{(L)} \lpar y^{(L-1)}\rpar
			= \half \left[ 1 + \tanh{\neurtanh} \right], \EQN nety
$$
and for
$$
\EQNalign{
	y_i^{(1)}	&= x_i, \cr
	L		&= 1, \dots ,M, \cr
	N^{(M)} &= 1, \cr
	N^{(1)} &= N.
}
$$
Note that the neural network function \f\ is a differentiable function of the
weights.

In order to train a neural network, the sets \sig\
and \bac\ must have a relative size approximately equal to the theoretical
signal to background ratio (in contrast with the case of binary decision tree
training if the scaling parameter $K$ is used),
because in practice, stochastic gradient descent
is substituted for classical gradient descent. In training with
stochastic gradient descent, a single event is chosen at random from
\sig\ or \bac, \f\ and its derivatives with respect to
the weights are calculated, and the weights $\weight[L,j,i]$ are
immediately rotated by a small amount toward the "downhill" direction of
the quadratic error function. This process is repeated for
all training events in random order and for many cycles.
In this way, the weights gradually and
smoothly move toward an optimal classification of the entire training with
much more modest computational demands than if classical gradient
descent were employed. In the latter case, however, because the order
of presentation of events does not matter (since weights are updated
only after presentation of the entire set), one can "scale" \bac\
 as necessary by multiplying each background event's contribution to the
quadratic error  function by the factor $K$ defined in \Eq{scale}. Training
however, requires much more computational time than the classical
gradient algorithm.

After successful training, the neural network function \f\
should take on values
greater than $\half$ for signal events and less than $\half$ for
background. In deriving a trigger from such a neural network,
one has the freedom to
specify the threshhold value $f(x_i) = \theta$ for an event to be
classified as signal. As $\theta \to 1$, the accepted events should
increase in signal purity and decrease in signal efficiency.
\section{Results \label{sect.results}}
In this section we present our results for the binary decision tree performance
in comparison with that of the neural network of \Ref{topnet} in obtaining a
ratio of signal to background events in the particular application of
top quark identification via the one-lepton channel at the Tevatron.

The training and testing event sets \sig\ and \bac\ are the same as those
used in \Ref{topnet} for which top quark production and the relevant
background are simulated by the Monte Carlo event generator PYTHIA\cite{pythia}
at $p{\bar p}$ center of mass energy 1.8~TeV. We consider
here a top quark mass of both 100 and 140~GeV. The $W$-boson
plus multijet background is also generated by PYTHIA from $q{\bar q} \to W g$
and $qg \to Wq$ subprocesses.
We reproduce the acceptance cuts applied to each event so generated; further
details of the simulation of training and testing sets are found
in \Ref{topnet}:
a)~one and only electron- or muon-type charged lepton of $p_T > 20$ GeV
and pseudorapidity $|\eta| < 3.0$,
b)~3 or more hadronic jets, each of energy 15~GeV and pseudorapidity
$|\eta| < 2.5$, for jet cone size  defined to be $\Delta r = 0.7$,
c)~total missing energy $\slashchar{E}_T > 20$ GeV, and
d)~lepton isolation such that the sum of hadronic energy within a cone of
size $\Delta r = 0.4$ centered about the lepton momentum is less than 3~GeV.

A number of parameters were used in the construction of the binary decision
trees to specify criteria for the classification and division of nodes.
Division was halted at a node if:
a)~it contained less than \minSB\ signal plus (scaled) background events
(1--50)\footnote*{For illustrative purposes, the value/range of each parameter
used in training the binary decision trees for the \mt=140~GeV signal are
indicated in parentheses.},
b)~it contained less than \threshSB\ total
events (10--10000) with a background event fraction higher
than \maxBSB\ (0.90--0.99),
c)~it possessed a sufficiently high signal event fraction
\minSSB\ (0.50--0.93),
or d)~all attempts at division would result in either the left or right branch
being less than \minLLR\ (0.0001) of the parent node.
A node thus terminated would be classified as a signal leaf if it
contained at least \minS\ (10--300) signal events,
and as a background leaf otherwise. The optimization process was controlled
by the choice of the temperature $T$ (0.0001--10) and the
minimum fractional reduction (0.05) required for continued iterations of
the conjugate gradient descent algorithm.

A group of binary decision trees was generated for each top quark mass by
varying the above parameters. A set of triggers
covering a range of signal to background
ratios and signal efficiencies was thus obtained. Each binary decision
tree was trained and tested on the same sets \sig\ and \bac\ used
in \Ref{topnet}. For the \mt=100~GeV, \sig\ consisted of
4500 points and \bac\ contained 5500, while for \mt=140~GeV,
\sig\ had 1500 points and \bac\ had 8500. Note that the relative sizes
of \sig\ and \bac\ approximated the respective signal to background ratios
of 0.77 and 0.19 for the 100 and 140~GeV top quark mass, respectively.
All binary decision trees were subsequently tested on sets of
2500 signal and 2500 background points.

Figure~1 shows the best results obtained for the binary decision
trees to recognize the top quark signal for \mt=140~GeV.  The efficiencies
plotted are simply the percentage of signal accepted by each trigger.
Plotted along\-side these data are a single point representing the
"severe" conventional cuts described
in \Ref{topnet} and the results obtained from the neural network trained
with the same \sig\ and \bac. The latter set of points was
produced by setting the neural network-derived
trigger threshold $\theta$ to $\{0, 0.1, 0.2, \dots, 1\}$. The binary
decision tree triggers more or less match their neural
network counterpart, though we were unable
to reproduce points with extremely high efficiency (but correspondingly low
signal to background ratios). A similar result is apparent from
Figure~2,  for \mt=100~GeV, which can be
attributed to the fact that despite the similarity of
\Eqs{approx} and \Ep{nety}, a neural network of at least 4 layers
partitions space in a very different fashion from the binary decision tree
(a neural network with only $M=3$~layers ---the minimum possible --- is
functionally identical to a binary decision tree with only two leaves.) Regions
 identified  by the neural network as signal are precisely those signal and
 background events for which $f(x_i) \ge \theta$. Examination
 of \Eq{nety} reveals that if $M \ge 4$,
the boundaries of these regions are complex curved surfaces arising from
inverting two or more recursive $\tanh$ functions. Regions classified as
signal by a binary decision tree, on the other hand, have hyperplanar
boundaries. Thus the neural network might fare better for
lower-efficiency cuts because the "ideal" partition in these cases would
enclose as
many signal events as possible within one contiguous region using
a smooth (non-planar) boundary.	In the region where the
trigger signal to background ratio is appreciably enhanced
over that of the parent
pre-cut distribution, the binary decision tree essentially matches the
performance of the neural network, and both gain substantial
improvement over the "severe" set of conventional cuts (see Figures~1 and 2).
For example, in one of the binary decision trees trained on  the \mt=140~GeV
data, the initial hyperplane alone managed to partition off a signal region
with the same efficiency (50\%) as the conventional cuts but with a signal
to background ratio of 1 instead of 0.65.

Although in this particular application the neural network and binary decision
tree obtained remarkably similar quantitative results, a few interesting
qualititative differences were observed as well.
As noted above, a neural network
of more than 3 layers divides up space in a quite complicated way. Even
supplied with the weights that define the neural network, it is in general
impossible to derive the boundaries of "signal" regions explicitly, which
makes error analysis of the experimental results from a neural
network-derived trigger quite difficult. By nature of its design, a
binary decision tree-derived trigger is comparatively transparent
in its operation, and fully equivalent to conventional
simplified cuts for error analysis
purposes. Furthermore, by examining the hyperplane normals of the trained
binary decision tree one may glean information regarding the relative
importance of the kinematic variables \xi\ for discrimination between
signal and background in each subregion of phase space.

The respective training phases of the neural network and the binary decision
tree differ greatly in the computational resources required.
Training of the binary decision trees required anywhere from 5\% down to
0.2\% of the cpu time used to train the neural networks of \Ref{topnet}, thus
bearing out the observation\cite{topnet} that
training time for a neural network should increase much more rapidly
with the addition of layers than that of a comparable binary decision tree.

The results for the neural network are more stable than those for the binary
decision tree in the sense that variations in the parameters that serve
to define the neural network, such as the temperature and the number of
hidden layers (\Ref{topnet}), do not appreciably change them.
The binary decision tree, on the other hand, produced a
wide range of results as parameters such
as the minimum signal percentage, \minSSB, were varied. In this application,
in fact, the neural network's performance was used as a benchmark, toward
which the binary decision trees were trained. The parameter \minSSB\
would initially be set equal to a given signal percentage attained by the
neural network, and other parameters would be varied in an attempt to match
or surpass the neural network's efficiency.

Finally, we note here that though not a limitation for this
particular application, an extremely small  signal to background ratio
would mean that to train a neural network, one would require a potentially
huge Monte Carlo-generated background training set \bac,
due to the use of stochastic gradient minimization. The binary decision tree,
in contrast, can scale the background set as necessary for any signal
to background ratio through $K$, so that \sig\ and \bac\  need only be large
enough to represent the theoretical distributions faithfully and in
sufficient detail.

The authors are grateful for much helpful discussion and participation by
D.A.~Dicus and G.F.~Giudice. We would like to acknowlege the aid of
B.~Merriman and P.~Burchard, whose work in binary decision trees
prompted our own investigations, and R.P.~Brent, whose algorithm is
used for hyperplane optimization. Finally, one of us (D.\,B-C)
expresses his appreciation for several insightful comments
on differences between binary decision trees and
neural networks made by E.I.~Levine.

\vfill
\supereject
\nosechead{References}
\ListReferences
\vfill\supereject
\section{Figure Captions \label{sect.captions}}
Figure 1: Results of training binary decision trees to recognize the top quark
signal assuming \mt=140~GeV; data for the corresponding neural network
 with the threshold
$\theta$ set (going from right to left on the graph)
to $\{0, 0.1, 0.2,\dots,1\}$ is included, as is a single point representing
the ``severe'' conventional cuts of \Ref{topnet}. The vertical
axis gives the expected signal to background ratio of the derived triggers
while the horizontal axis displays the signal efficiency (the percentage
of signal accepted by each trigger).

Figure 2: Results for the binary decision trees, neural network,
and set of conventional cuts
 for a top quark mass of \mt=100~GeV. Note that the leftmost data point
 for the neural network was omitted because of insufficient statistics.
\end